# Possible Magnetic Rotational Bands in $^{84}$Rb

Shuifa Shen[1 2 3 4 5*], Jiejie Shen[6], Guangbing Han[7], Shuxian Wen[8], Yupeng Yan[2 3], Haibing Jiang[5]
Zhixiong He[5], Yaodong Song[5], Weiwei Gao[5], Jieqing Lan[5]

([1]*Institute of Nuclear Energy Safety Technology, Chinese Academy of Sciences, Anhui, Hefei 230031, People's Republic of China*)

([2]*School of Physics, Suranaree University of Technology, Nakhon Ratchasima 30000, Thailand*)

([3]*Thailand Center of Excellence in Physics (ThEP), Commission on Higher Education, 328 Si Ayutthaya Road, Ratchathewi, Bangkok 10400, Thailand*)

([4]*Key Laboratory of High Precision Nuclear Spectroscopy, Institute of Modern Physics, Chinese Academy of Sciences, Gansu, Lanzhou 730000, People's Republic of China*)

([5]*School of Electronic, Electrical Engineering and Physics, Fujian University of Technology, Fujian, Fuzhou 350118, People's Republic of China*)

([6]*School of Medicine, Hangzhou Normal University, Zhejiang, Hangzhou 310036, People's Republic of China*)

([7]*School of Physics, Shandong University, Shandong, Jinan 250100, People's Republic of China*)

([8]*China Institute of Atomic Energy, P. O. Box 275(10), Beijing 102413, People's Republic of China*)

Abstract: High-spin states in $^{84}$Rb are studied by using the $^{70}$Zn($^{18}$O, p3n)$^{84}$Rb reaction at a beam energy of 75 MeV. Three high-lying negative-parity bands are established, whose level spacings are very regular, i.e., there is no signature splitting. The dipole character of the transitions of these three bands is assigned by the γ-γ directional correlations of oriented states (DCO) intensity ratios and the multipolarity M1 is suggested by analogy with multiparticle excitations in neighboring nuclei. Strong M1 and weak or no E2 transitions are observed. All these characteristic features show they are magnetic rotational bands.

Keywords: In-beam γ-spectroscopy; Magnetic dipole band; B(M1)/B(E2).

## 1. Introduction

Magnetic rotation, a novel kind of nuclear rotation, has attracted a great interest in recent years. The levels of rotational bands are linked by strong magnetic dipole (M1) transitions whereas crossover electric quadrupole (E2) transitions are very weak. The ratios of the transition probabilities B(M1)/B(E2) are large. This magnetic character of the rotation is demonstrated by the ratios of transition probability B(M1)/B(E2) from each level in the band.

*E-mail address: shuifa.shen@inest.cas.cn

The experimental evidence for magnetic rotational bands includes the presence of a greater intensity of the ΔI=1 M1 transitions between neighboring levels within one band in some nuclei with small deformation. These M1 transitions are different from those M1 transitions usually observed in high-spin states. At first, the energies of these transitions are very regular, that is, there is no signature splitting, which is very similar to high-K rotational bands in well deformed nuclei. Secondly, their magnetic dipole reduced transition probability B(M1) values are greatly enhanced, can up to several $\mu_N^2$ units. Therefore, the magnetic dipole and electric quadrupole reduced transition probability ratio B(M1)/B(E2) is very large. The E2 transitions within the band are very weak or cannot be observed, which is different from a high-K band in a well deformed nucleus. This indicates that the deformation corresponding to this band is very small. Moreover, the ratio of the dynamic moment of inertia $J^{(2)}$ to electric quadrupole reduced transition probability B(E2) is larger than that of normally deformed bands and superdeformed bands. It can be as large as 10 times more.

In recent years, the study of magnetic rotational bands has been given great attention, either in theories or in experiments. At first, in previous years, magnetic rotational bands have been found in some nuclei in the Pb region, e.g., in $^{199}$Pb and $^{200}$Pb etc., in $^{139}$Sm around the A~140 mass region, in $^{110}$Cd and $^{105}$Sn around the A~110 region, and in $^{82}$Rb [1] and $^{84}$Rb [2-4] around the A~80 mass region etc. From the theoretical side, shell model calculations as well as relativistic mean – field (RMF) descriptions of the shears band mechanism in $^{84}$Rb were accomplished in this mass region more than ten years ago [3, 5], and its adjacent nucleus $^{82}$Rb was also studied using the *complex* Excited Vampir approach [6]. The present work focuses on the magnetic dipole bands in $^{84}$Rb and complements our preceding publication [7].

## 2. Assignment of magnetic dipole bands in $^{84}$Rb

High-spin states in $^{84}$Rb were studied by the heavy-ion fusion-evaporation reaction $^{70}$Zn($^{18}$O, p3n)$^{84}$Rb using an $^{18}$O beam provided by the HI-13 tandem accelerator at the China Institute of Atomic Energy (CIAE). Details of the experimental procedure and results were published in Ref. [7], where the negative-parity bands were extended greatly from the previous ($6^-$) up to the highest ($17^-$) spins, and the spins and parities of these levels were tentatively assigned based on γ-γ directional correlations of oriented states (DCO) intensity ratios [8] and previous works. The γ-γ-coincidence data were analyzed with the Radware software package [9]. For convenience in discussing γ-γ coincidence relations in detail below and for completeness, here the level scheme is also given, as shown in Fig. 1.

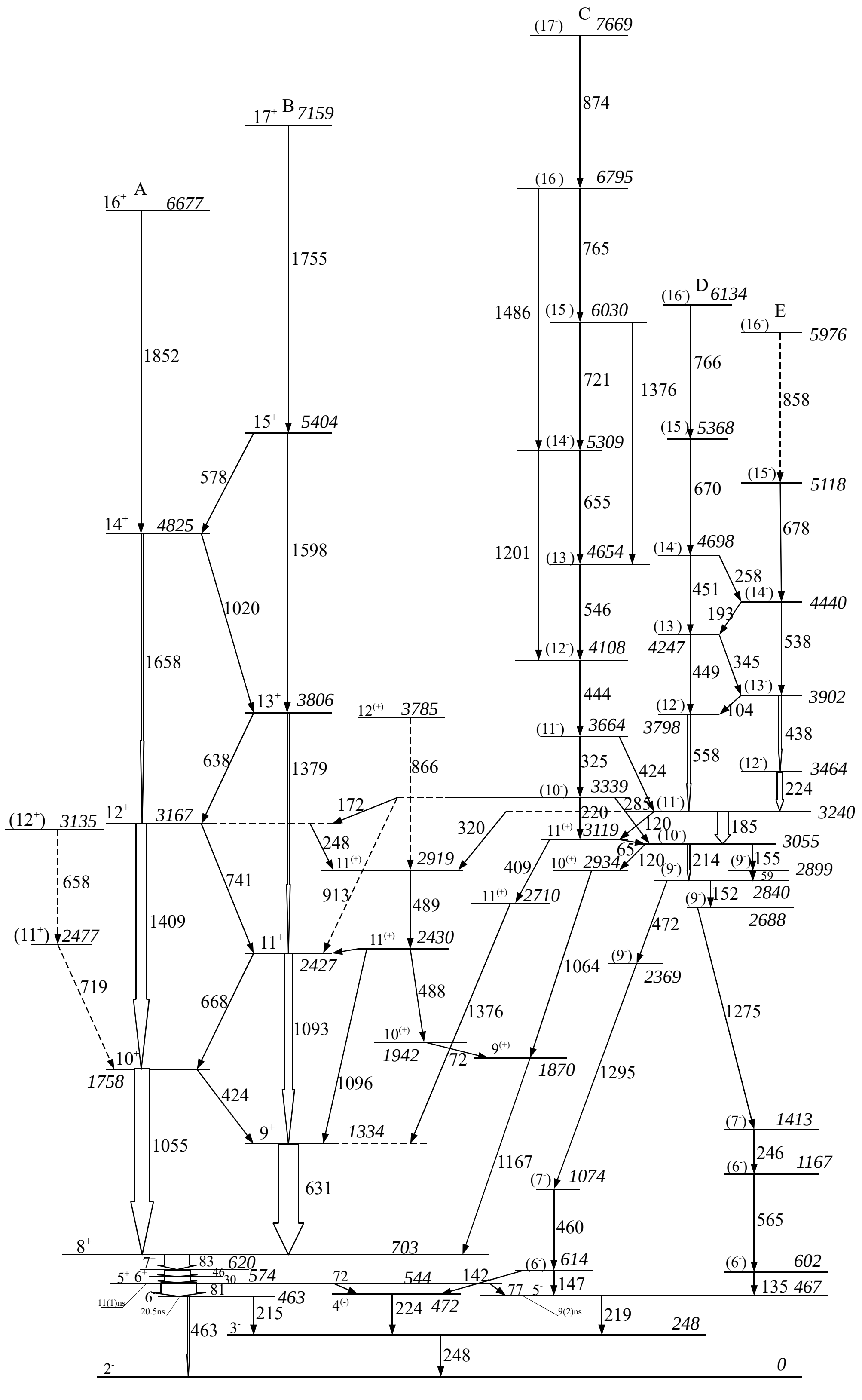

Fig. 1. A partial level scheme of $^{84}$Rb proposed in the present work with the width of the γ transitions proportional to their measured intensities. The transition energies are given in kiloelectron volts.

In the following we focus on the assignment of the cascade relationships relevant to these three bands

(denoted as bands C, D and E in Fig. 1): In the spectrum gated by the 83 keV γ-ray, besides the new found γ rays which are used to extend the positive-parity yrast band, there are still some very strong γ rays at 185, 325, 409, 488 and 719 keV. Fig. 2 shows these γ rays clearly. Fig. 3 shows the spectrum gated by 185 keV.

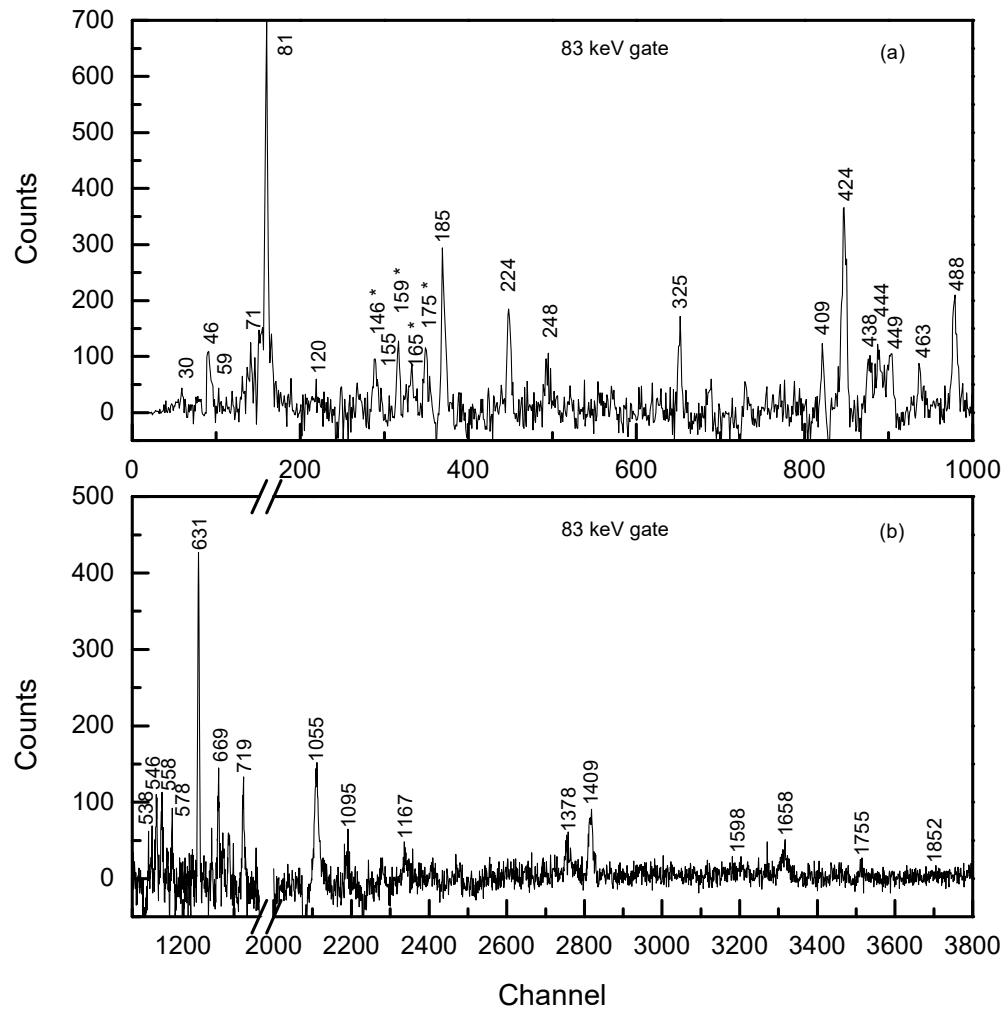

Fig. 2. Gamma-ray spectrum created by gating on the 83 keV γ-ray. (a) The low-energy region. (b) The high-energy region. γ rays marked with an asterisk do not belong to $^{84}$Rb.

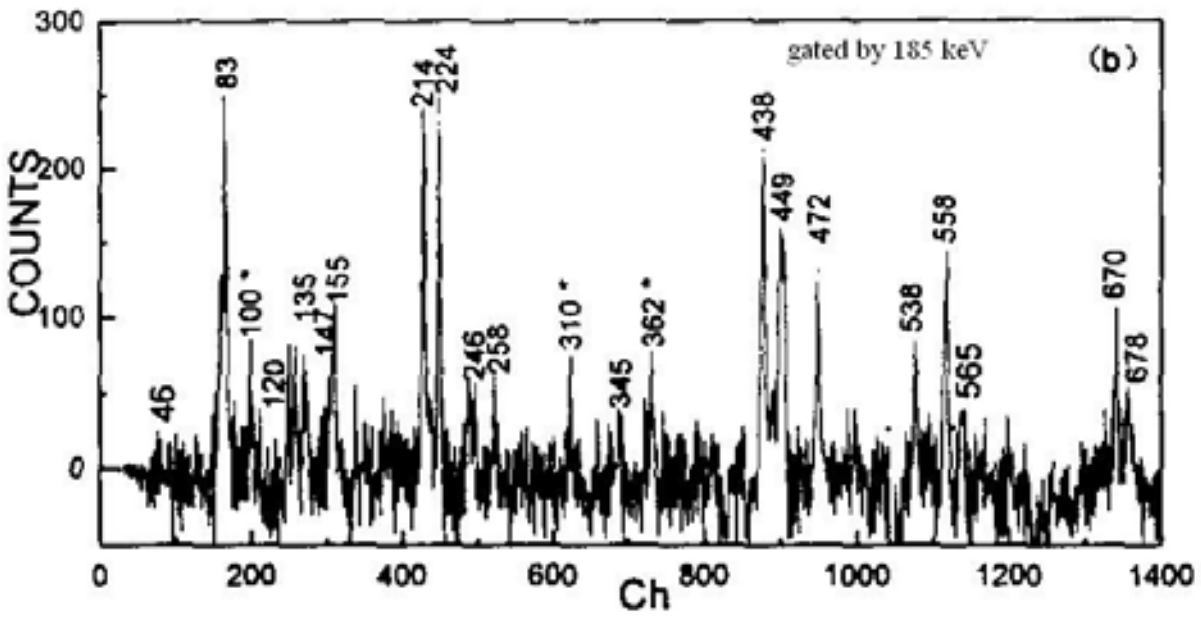

Fig. 3. Similar to Fig. 2, but for the spectrum gated by the 185 keV γ-ray.

One of the most important things that attracts our attention is some relatively strong γ rays at 224, 438, 449, 538, 670 keV and others. In the spectrum gated by the 83 keV γ-ray these γ rays are also observed, and furthermore the peaks are not small. Therefore, the cascade relationship between the 83 keV γ-ray and the 224, 438, 558, 449 and 538 keV γ rays has been given a preliminary assignment, as also has the cascade relationship between the 185 keV γ-ray and them. From the intensity relation shown in Fig. 2 it can be deduced that the 185, 224 and 438 keV γ rays have a direct cascade relationship. This is also proven by the spectra gated by the 224 and 438 keV γ rays. At the same time, the 538, 451, 345, 670 and 678 keV γ rays can be found from these two spectra, but the 558 keV γ transition can't be found. Hence, it can be seen that the 558 keV γ-ray does not have the cascade relationship with them. However, from the spectrum gated by the 558 keV γ-ray, the 83, 46, 185 and 631 keV γ rays can be clearly found, and also the very strong 449, 538, 670 and 678 keV γ rays, thus showing that the 558 keV γ-ray belongs to $^{84}$Rb. When gated by the 449 keV γ-ray, we observe the 558 keV γ-ray as expected. Comparing the spectra gated by the 538, 670 and 678 keV γ rays, it can be determined that the cascade relationship of the 538 and 449 keV γ rays is attributable to $^{78}$Kr, whereas the appearance of the 438 keV γ-ray in the spectrum gated by 449 keV γ-ray is attributed to the cascade relationship of the 451 and 438 keV γ rays (the 449 keV γ-ray is close to the 451 keV γ-ray). So, after careful comparison and identification, the cascade relationship is built above the 185 keV γ-ray as shown in Fig. 1, but how about the position relation between these cascades and this 83 keV γ-ray? From Fig. 3 it is known that 185 keV γ-ray has no cascade relationship with the 631 keV γ-ray, but has a cascade relationship with the 224, 558 keV γ rays and others. After comparison of the spectra gated by the 185 and 83 keV γ rays, it is found that they share the relatively high-energy 1167 and 1064 keV γ rays and the low-energy 120 keV γ-ray, thus supporting a preliminarily assignment that they connect the 83 and 185 keV γ rays. Then the analysis of the spectra gated by the 1167, 1064 and 120 keV γ rays substantiates this viewpoint. Fig. 3 also has the remaining strong 214, 472, 155 keV γ lines and others. The 152, 472, 1275, 1295 keV γ rays and others are observed in the spectrum gated by the 214 keV γ ray, but it has no cascade relationship with the 83 and 631 keV γ transitions. The 214 and 472 keV γ transitions can also be observed clearly in the spectra gated by the 224, 438, 558 keV γ transitions and others. It can thus be seen that the 214 keV γ-ray is placed under the 185 keV γ-ray. Because the peak of the 472 keV γ transition is relatively clearly visible in the spectrum gated by the 214 keV γ transition, so the 472 keV γ transition is placed under the 214 keV γ transition. The spectrum gated by the 472 keV γ transition also supports this viewpoint. Through the comparison of the spectra gated by the 185, 214, 472 keV γ transitions and others, the cascade of the 152, 1275, 246 and 565 keV γ transitions is built under the 214 keV γ transition, and that of the 1295, 460 keV γ transitions and others under the 472 keV γ transition. In addition, both the 155 and 59 keV γ transitions are clearly observed in the spectra gated by the 185 and 472 keV γ transitions, whereas they are not clear in the spectrum gated by the 214 keV γ transition. Because 155+59=214, so the 155 and 59 keV γ transitions are suggested to be placed into the level scheme as shown in Fig. 1.

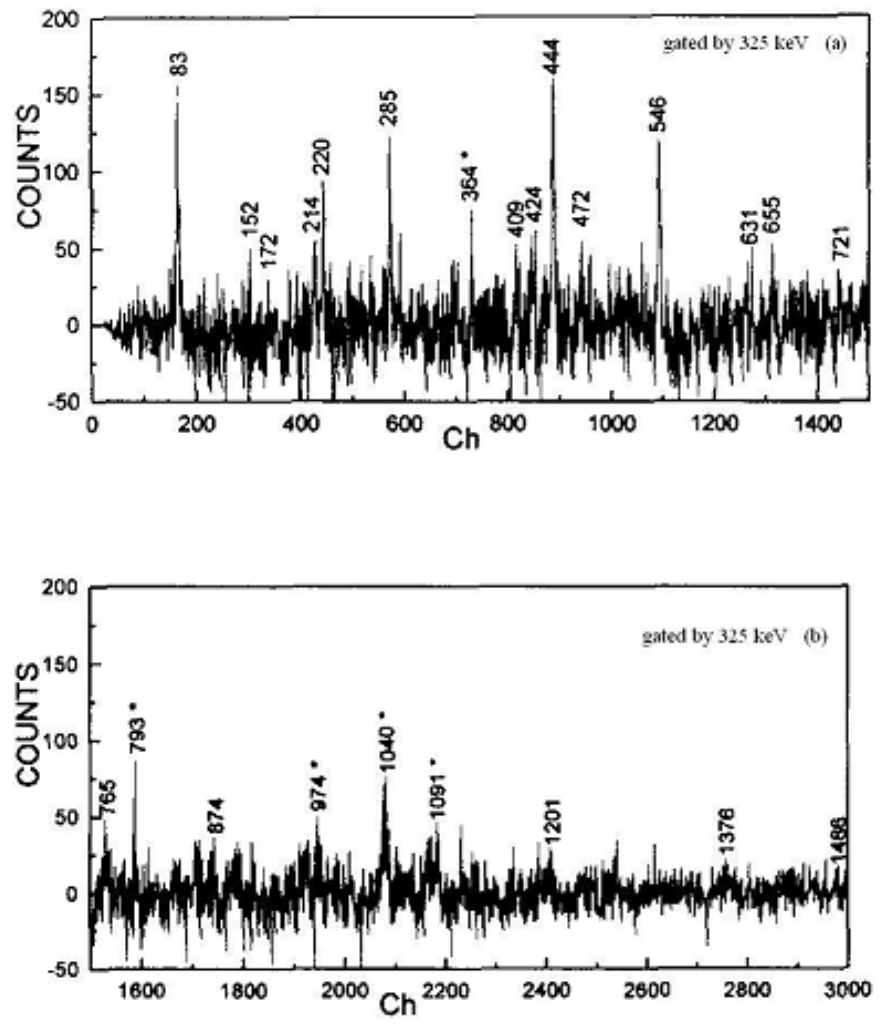

Fig. 4. Similar to Fig. 2, but for the spectrum gated by the 325 keV γ-ray.

Fig. 4 gives the spectrum gated by the 325 keV γ transition from which very strong γ transitions at 83, 444, 546 keV and others are observed. After comparison of the spectra gated on the 220, 444, 546, 655 keV γ transitions and others one by one, their transition sequence is assigned as is shown as band C in Fig. 1. It should be noticed that in their gated spectra both the 214 and 409 keV γ transitions can be observed. Except for the spectra gated by the 220 and 325 keV γ transitions, the 185 keV γ transition is also observed in other spectra. After comparison of the spectra gated by the 325 and 214 keV γ transitions, the 285 keV γ transition is assigned between them, and the 65 keV γ transition between the 214 and 220 keV γ transitions. In addition, in the spectrum gated by the 325 keV γ transition a clear 1093 keV γ transition is observed, so it is speculated that maybe a 913 keV γ transition exists between them, but in the 325 keV γ transition gated spectrum it is not obvious. Although the spectrum gated by the 409 keV γ transition is more complicated, it has relatively clear 83, 631, 1376, 220, 224, 325 keV γ transitions and others, thus enabling the assignment of the cascade relationship among the 409, 1376, 631 and 83 keV γ transitions, and the cascade relationship of the 409 keV γ transition with the 224, 438, 558 keV γ transitions and others through the 120 keV γ transition.

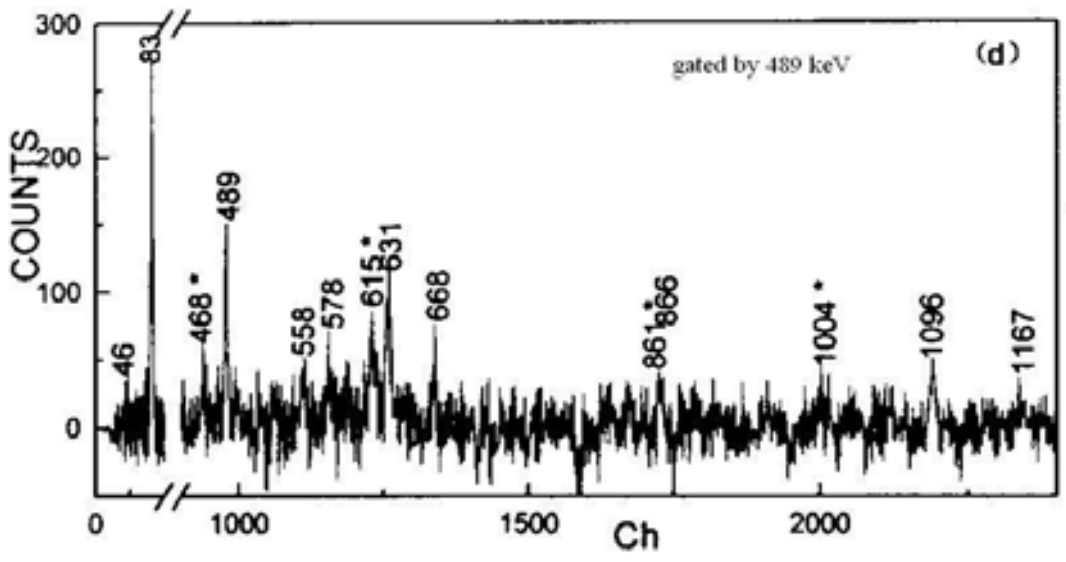

Fig. 5. Similar to Fig. 2, but for the spectrum gated by the 489 keV γ-ray.

There are still relatively strong 489 and 719 keV γ transitions in Fig. 2. Fig. 5 shows the spectrum gated by the 489 keV γ transition, the 83, 488, 631 and 1096 keV γ transitions can be clearly observed. From the intensity it can be seen that the 489 keV transition has a direct cascade relationship with the 488 and 1096 keV γ transitions, respectively. The cascade relationship of 489, 1096 and 631 keV γ transitions also can be seen in the spectrum gated by the 631 keV γ transition. The 1167 keV γ transition can also be observed in Fig. 5. After careful inspection, the 72 keV γ transition is found which locates between the 1167 and 488 keV γ transitions. In addition, there is a relatively clear γ transition at 668 keV, so it can be predicted that there is a 3 keV γ transition between the 489 and 668 keV γ transitions. It is proven by the observed 489 keV γ transition in both the spectra gated by the 424 and 1055 keV γ transitions. The 578 keV γ transition has a relatively strong intensity in the spectrum gated by the 489 keV γ transition because there exists a 248 keV γ transition between the 489 and 1658 keV γ transitions. The relatively weak 224, 438, 558 keV γ transitions and others can also be observed in Fig. 5. This is because the 489 keV γ transition has a cascade relationship with them through the 320 keV γ transition.

The remaining relatively strong transition is the 719 keV γ-ray. After gating on it we find its main component belongs to $^{84}$Sr, but there are relatively clear 83 and 46 keV γ transitions in its gated spectrum, which tell us that there exists a 719 keV γ transition in $^{84}$Rb. After comparing with the 83 keV γ-ray gated spectrum we assign their transition sequence.

In the present work we concentrate on the most interesting feature, which is the sudden development of a regular magnetic dipole band at excitation energy around 3 MeV (see Fig. 1). Note that the difference in the first M1 band (denoted as Band C) between our work and that of Schnare et al. [2-4] is one spin unit.

Among the three bands (denoted as bands C, D and E), the band C magnetic-rotation assignment is discussed first. It consists of seven strong M1 transitions of 325, 444, 546, 655, 721, 765, and 874 keV on top of the $I^{\pi}=(10^{-})$ level at $E_x=3.339$ MeV. The dipole character of the transitions is proven by the γ-γ directional correlations of oriented states (DCO) intensity ratios and the M1 multipolarity is suggested by analogy with multiparticle excitations in neighboring nuclei. From its Routhian it can be found that there is no signature splitting. It can be observed from Fig. 2 and Fig. 4 that the 325, 444, 546, 655, 721 keV and other M1 transitions are much stronger than the 1201, 1376, 1486 keV and other E2 γ transitions. The ratios of reduced transition probabilities B(M1)/B(E2) are extracted. The error of these ratios is relatively large because the E2 transitions are very weak, but the trend of the ratios can be seen. The above experimental results show that this band C has the magnetic-rotation characteristics. No E2 crossover transition is observed in bands D and E, which are also identified as magnetic rotational bands. It should be pointed out here that any two bands can never have the same band head. Since we tentatively assign the ($11^-$) 3240 keV level as the band head of band E, the band head of band D can only be the ($12^-$) 3798 keV level.

In the present work, the features of the three dipole bands shown in Fig. 1 are compared with the general criteria for a magnetic-rotation band in which the level energies (E) versus the spins (*I*) follow the pattern

$$E - E_0 = A_0(I - I_0)^2, \qquad (1)$$

where $E_0$ and $I_0$ are the energy and spin of the band head, respectively, and $A_0$ is a constant. We plot $E-E_0$ versus $(I-I_0)^2$ in Fig. 6a, 6b and 6c for bands C, D, and E, respectively. The solid lines in the figure are the fits to the data using the relation in Eq. (1). The good agreement of the data with the fitted lines, as shown in Fig. 6, indicates that bands C, D and E all follow the relation in Eq. (1). The dynamic moments of inertia $J^{(2)}$ obtained for these three bands with

$$J^2(I) \approx \frac{2\hbar^2}{\Delta E_\gamma(I)} = \frac{2\hbar^2}{E_\gamma(I+1) - E_\gamma(I)}, \qquad (2)$$

where $E_\gamma(I) \equiv E_\gamma(I \rightarrow I-1)$, are within the typical range of $J^{(2)}$~10-25$\hbar^2$MeV$^{-1}$ for a magnetic rotational band. Therefore, one may conclude that each of the above three bands is most likely to be a magnetic rotational band.

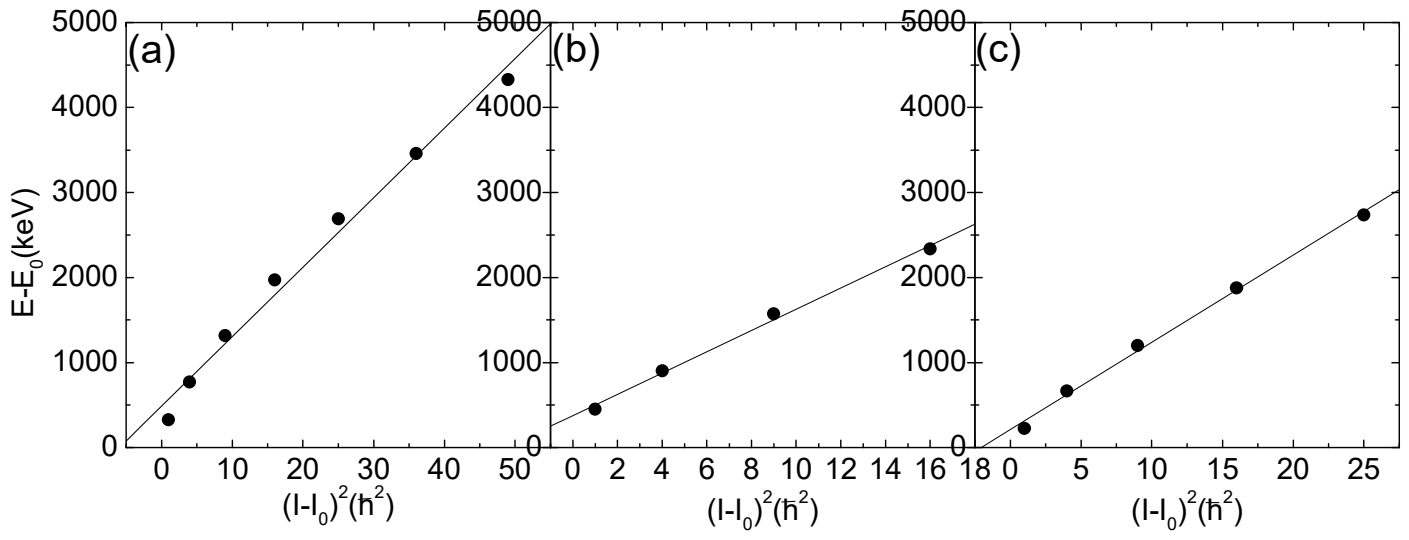

Fig. 6. Relative energy (E) versus spin (*I*) curve for (a) band C, (b) band D, and (c) band E, respectively. $E_0$ and $I_0$ are the band-head energy and spin, respectively. The fitted curves are shown by the solid lines (see text for details).

## 3. Conclusion

High-spin states in $^{84}$Rb have been populated in the reaction $^{18}O+^{70}Zn$ at a beam energy of 75 MeV. By analyzing the γ-γ coincidence data, three negative-parity M1 sequences are observed, which show the characteristic features of magnetic rotation such as regular level spacings and large reduced transition probability ratios, of the magnetic dipole (M1) transitions to the electric quadrupole (E2) transitions, B(M1)/B(E2). The dynamic moments of inertia $J^{(2)}$ obtained for these three bands are also within the typical range of $J^{(2)}$~10-25$\hbar^2$MeV$^{-1}$ for magnetic rotational bands. Therefore, one may conclude that each of these three bands is most likely to be of magnetic rotational character.